\documentclass[doublecol]{epl2} 
\usepackage{amsmath,amssymb}

\title{Dynamical bistability in the driven circuit QED}
\shorttitle{Dynamical bistability in the driven circuit QED} %Insert here a short version of the title if it exceeds 70 characters

\author{V. Peano\inst{1,3}  and M. Thorwart\inst{2,3}}
\shortauthor{V. Peano \etal}

\institute{                    
  \inst{1} Physikalisches Institut, Albert-Ludwigs-Universit\"at Freiburg,
79104 Freiburg, Germany\\
  \inst{2} Freiburg Institute for Advanced Studies (FRIAS), Albert-Ludwigs-Universit\"at Freiburg,
79104 Freiburg, Germany \\
\inst{3}Institut f\"ur 
Theoretische Physik, Heinrich-Heine-Universit\"at D\"usseldorf,
40225  D\"usseldorf, Germany 
}

\abstract{We show that the nonlinear response of a driven 
circuit 
quantum electrodynamics setup displays antiresonant multiphoton 
transitions, as recently observed in a transmon qubit device. 
By including photon leaking, we explain the lineshape 
by  a perturbative and a semiclassical analysis.
We derive a bistable semiclassical quasienergy surface whose lowest quasienergy eigenstate is squeezed, allowing to define a squeezing-dependent 
local effective temperature. We study the escape dynamics 
out of the metastable state and 
find signatures of dynamical tunneling, 
similar as for the quantum Duffing oscillator.
}

\begin{document}

\maketitle

\section{Introduction}
One of the nontrivial fundamental models of quantum physics is the Jaynes-Cummings (JC) model 
\cite{jaynescummings}. It was introduced to describe the interaction of a two-level atom and a single 
quantized electromagnetic field mode. Being sufficiently simple, its dynamics is 
very rich though, including Rabi oscillations, collapse and revival phenomena, 
squeezing, entanglement, Schr\"odinger cat and Fock states, and photon 
antibunching~\cite{Larson}. 
Beyond quantum optical set-ups, it is applicable to 
many situations of nanocircuit quantum electrodynamics (QED), such as 
Cooper pair boxes~\cite{Wallraff}, superconducting 
flux qubits~\cite{Chiorescu}, Josephson junctions~\cite{Hatakenaka}, and semiconductor quantum dots~\cite{Winger}.  
In particular, the latter setups allow to explore the regime of 
strong coupling and nonlinear response. 

Recently, unique nonlinear features have been detected in the transmitted 
heterodyne signal of a superconducting transmon qubit device~\cite{Bishop}. For weak driving, the two well-known 
vacuum Rabi resonances reflect transitions
between the JC groundstate and the first/the second excited state of the 
undriven JC spectrum.
Their difference in energy is $2\hbar g$, where $g$ 
is the interaction strength of the qubit and the harmonic mode. 
For increasing driving, 
each vacuum Rabi peak  supersplits  into additional (anti-)resonances with the
characteristic $\sqrt{n}$ spacing. The measurements have been corroborated with
accurate numerical simulations based on a Markovian master equation of Lindblad form ~\cite{Bishop}. 
For the particular case of the one-photon resonance of the JC groundstate and the antisymmetric 
superposition of transmon and photon excitation, an effective two-level model composed of both transmon and 
cavity degrees of freedom has been established. The ensuing Bloch equations have been solved analytically, yielding a closed 
expression for the heterodyne amplitude. For higher multiphoton resonances, the numerical solution of the 
master equation was compared to the experimental data, with superb agreement. However, 
the underlying nontrivial competition of driving and dissipation has not been discussed.

In this Letter, we provide a complete physical picture for the dissipative dynamics of 
the Purcell-limited nonlinear response 
 of the driven JC model in terms of quantum multiphoton (anti-)resonances.  The underlying 
competition of driving and dissipation is revealed by perturbative 
arguments in the rotating frame in presence of photon leaking. We show explicitly that the 
 lineshape of the multiphoton resonance in general is determined by the ratio of the Rabi frequency 
and the dissipation strength, allowing for a direct experimental control. 
In addition to the general analysis of the lineshapes of the multiphoton resonances, which 
were discussed in Ref.\ \cite{Bishop} based on the numerical solution of the master equation only, we
consider the semiclassical limit of the driven JC model. In the regime of strong driving, which cannot be accessed 
perturbatively,  we derive a novel semiclassical quasienergy surface which is {\em bistable\/}. 
We show that its lowest quasienergy eigenstate is an amplitude squeezed state and 
displays large out-of-phase oscillations. We consider the dissipative dynamics 
on the quasienergy surface at zero temperature. It is known that dissipation 
in driven nonlinear oscillators introduces nontrivial effects, such as 
quantum activation~\cite{Dykman89,Dykman06,Dykman07} and dynamic resonant tunneling~\cite{Peano1,Peano2,Peano3}. We 
 show that for the case of the dissipative driven JC model, the  
lowest quasienergy eigenstate is significantly populated at a multiphoton (anti)resonance, and is metastable 
away from resonance. The semiclassical analysis allows for an intuitive 
picture of dynamic bistability and dynamic resonant tunneling for the driven 
dissipative JC model, which has not been developed before.
Furthermore, we reveal 
deep analogies with the quantum Duffing oscillator~\cite{Dykman80,Dmitriev,Dykman89,Peano1,Dykman,Peano2,Peano3,Dykman07}. 

\section{The model}
We start from a harmonic oscillator with frequency $\omega_{\rm r}$ which is resonantly coupled with strength $g$ 
to a qubit of the same frequency $\omega_{\rm r}$ and which is driven with frequency $\omega_{\rm ex}$ and field strength $f$.  
 In the frame rotating with $\omega_{\rm ex}$ and 
for $\delta \omega \equiv \omega_{\rm r}-\omega_{\rm ex}, g, f\ll \omega_{\rm r}$, we perform 
 a rotating-wave approximation and obtain the Hamiltonian of the driven JC model   ($\hbar=k_B=1$)
\begin{equation}\label{hamrwa}
 H=\delta \omega \left( a^\dagger a+\frac{\sigma_z}{2}+\frac{1}{2}\right) +\frac{g}{2} \left(a^\dagger \sigma_- +a\sigma_+\right)
  + \frac{f}{2} \left(a^\dagger+a\right) \, ,
\end{equation}
with $\sigma_\pm=\sigma_x\pm i\sigma_y$. 
Here, $\sigma_j$ are the Pauli matrices. 
The undriven JC model has the quasienergies $\varepsilon_{0}=-\delta\omega/2\,,\quad \varepsilon_{n, \pm}=(n-1/2)\delta \omega \pm g\sqrt{n}$ and the quasienergy 
states $|\phi_0\rangle=|0, g\rangle,\quad |\phi_{n\, \pm}\rangle=(|n-1, e\rangle\pm|n, g\rangle)/\sqrt{2}$. 
 We will refer to latter as $n$-photon (dressed) states with two spin directions $\pm$. 
For $f\ne 0$, avoided crossings of the quasienergy levels arise,  which correspond to $N$-photon transitions 
 at $\delta\omega=\pm g/\sqrt{N}$.  To have well separated resonances, 
we consider the regime $g \gg f$. Around the resonance, the zero-photon state $|\phi_0\rangle$ 
and the $N$-photon dressed state $|\phi_{N\pm}\rangle$ display Rabi oscillations with 
the Rabi frequency $\Omega_N \propto f^N/g^{N-1}$. 

Due to the suppression of $1/f-$noise, transmon qubits are  Purcell-limited in the resonant regime considered here~\cite{Houck}. For low temperatures 
 $T\ll \omega_{\rm ex}$, we incorporate Purcell dissipation by means of the 
simple Lindblad master equation 
\begin{equation}\label{me}
 \dot{\rho}=-i[H,\rho]+\frac{\kappa}{2}\left([a\rho,a^\dagger]+[a,\rho a^\dagger]\right)\, ,
 \end{equation}
for the reduced density operator $\rho$ of the coupled qubit-oscillator system. 
%%%%%%%%%%%%%%%%
%%%%%%%%%%%%%%%%%%
Then, only photon  leaking from the system into the bath is possible. Hence, in absence of multiphoton transitions, 
$|\phi_0\rangle$ is dominantly populated in the stationary state.  In contrast, at a multiphoton resonance, the stationary state is generated by  coherent driving to
 the $N$-photon state and a subsequent relaxation  
via all the intermediate $n$-photon states ($n\le N$) to the $0$-photon state due to photon leaking. Eventually, 
this nontrivial interplay generates a stationary mixture of all $n$-photon states. In particular, 
each different choice of model parameters influences the intermediate dissipative transitions, which, in turn, 
determine the asymptotic populations of the stationary state.  

We are interested in the nonlinear response characterized by 
the steady-state intracavity field $\langle a  \rangle = {\rm tr}(\varrho  a) = A e^{i\varphi}
 = \sum_{\alpha \beta}\varrho_{\alpha\beta} a_{\beta\alpha}$, where $\alpha,\beta$ refer to 
the quasienergy eigenstates basis. 
We discuss the case $\delta \omega > 0$, the opposite follows from 
$|\phi_{n\pm} \rangle \to |\phi_{n\mp} \rangle$, $f \to -f$ and $\varphi \to -\varphi+\pi$.   
The modulus $A$ is related to the  transmitted amplitude 
$A_{\rm tr}\propto A$ and intensity $ I_{\rm tr}\propto A^2$ 
of the microwave input signal at frequency $\omega_{\rm ex}$~\cite{Bishop}. In the rotating frame, $\langle a\rangle< 0$ ($\varphi=\pi$) 
corresponds to an oscillation out of phase with respect to the drive.
\begin{figure}
\begin{center}
\includegraphics[width=80mm,keepaspectratio=true]{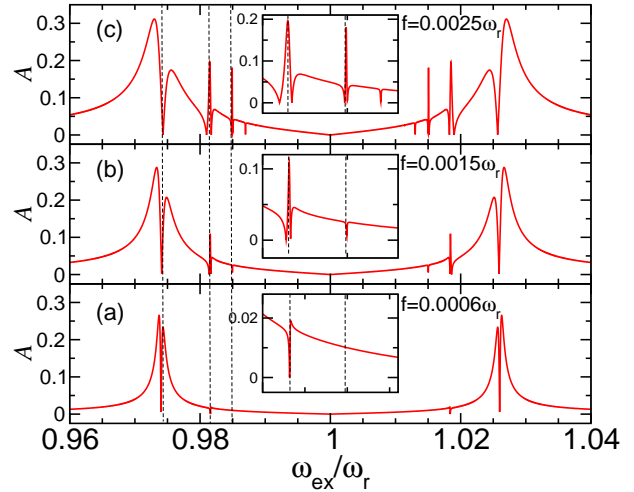}
\caption{\label{fig.1}Amplitude $A$ as a function of the driving frequency $\omega_{ex}$ for $f=0.0006 \omega_{\rm r}$ (a), $f=0.0015 \omega_{\rm r}$ (b) and $f=0.0025 \omega_{\rm r}$ (c). Insets: zooms to the corresponding (anti-)resonances marked by the dashed lines. Moreover, $g=0.026 \omega_{\rm r}, \kappa=5 \times 10^{-5}\omega_{\rm r}$. 
A realistic value for $\omega_{\rm r}/(2\pi)$ is $7$ GHz~\cite{Bishop,Houck}.  
}
\end{center}
\end{figure}
\section{Numerical results}
\begin{figure}
\begin{center}
\includegraphics[width=88mm,keepaspectratio=true]{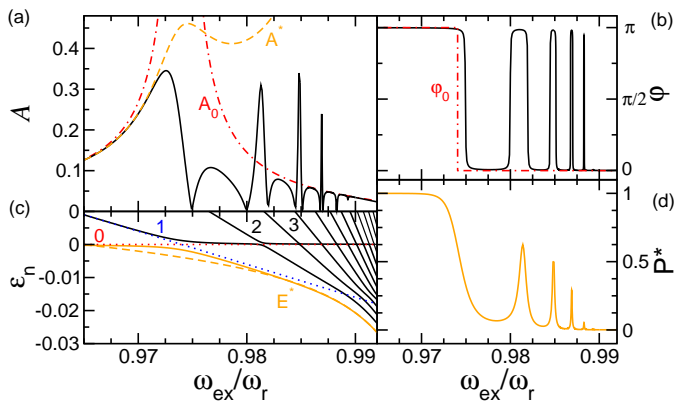}
\caption{\label{fig.2}Nonlinear response of the driven JC model: (a) amplitude, (b) phase,  (c) quasienergies, and (d) population of the lowest quasienergy state $|\psi^*\rangle$ for $g=0.026 \omega_{\rm r}, f=0.004 \omega_{\rm r}, \kappa= 10^{-4} \omega_{\rm r}$. Dashed-dotted red line in (a,b): lowest-order result for non-resonant approximation. Dashed orange line in (a): $A^*=|\langle \psi^* | a | \psi^* \rangle|$; and 
in (c): semiclassical result 
eq.~(\ref{estar}) 
for the lowest quasienergy $E^*$. 
}
\end{center}
\end{figure}

The nonlinear response, in the first instance obtained from a numerical solution of Eq.\ (\ref{me}), is shown in Fig.~\ref{fig.1} for the parameters in the regime  of Ref.\ \cite{Bishop}.  The drive induces a splitting of the vacuum Rabi resonance and 
produces two families of peaks which are symmetric with respect to $\omega_{ex}=\omega_{\rm r}$ and which are associated to the $\pm$ quasienergy states.  In each family, (anti)resonances occur which correspond to multiphoton transitions and which are associated to the avoided crossings 
of quasienergy levels, see Fig.~\ref{fig.2}c. We note that no antiresonances occur in the photon number $\langle a^\dagger a\rangle$~\cite{Chough}. 

For weak driving (Fig.~\ref{fig.1}a), only the $1$-photon antiresonance is well pronounced. It can be described~\cite{Bishop,Peano3} by a  model involving the $0$- and the $1$-photon state.
The generic mechanism is that, at resonance, they are equally populated and oscillate with opposite phase yielding zero response. Slightly away from resonance, one of the two states is more populated and a finite response arises. Far away from the resonance, the response again approaches zero. Overall, the lineshape of an antiresonance arises. The antiresonance around $\omega_{ex}\approx 0.98 \omega_{\rm r}$  corresponds to a $2$-photon process  
 (Fig.~\ref{fig.1}a). This feature also follows from a two state description. The shape is different from the 
$1$-photon antiresonance, due to the background contribution of nonresonant $1$-photon mixing processes~\cite{Dykman}.

For increasing driving, unexpected features arise. The multiphoton antiresonances turn into resonances, see  Fig.~\ref{fig.1}b and c. Associated to the behavior of $A$ are jumps 
in the phase $\varphi$, see Fig.~\ref{fig.2}b. This already suggests that two quasienergy states - one oscillating in and one out of phase - are alternately populated. 
Interestingly,  the population (Fig.~\ref{fig.2}d) of the state $|\psi^*\rangle$ with
 lowest quasienergy  shows peaks at the resonance frequencies. In fact,  as we will show below,  $|\psi^*\rangle$ is 
localized  in the bottom of a well of a bistable quasienergy surface. $|\psi^*\rangle$ is metastable since the bath induces
 transitions to higher quasienergies states and an escape is always possible, even at zero temperature.  This feature
 has also been reported for the quantum Duffing oscillator~\cite{Peano1,Peano2,Peano3}.

\section{Perturbative approach}
These observations are further substantiated by perturbative arguments, which show how the asymptotic populations 
of the stationary state are determined by the various rates of the intermediate dissipative transition 
processes. Here, we assume that the dressed states  $|\phi_{n\pm}\rangle$ ($n\neq N$) are approximate eigenstates. 
In the next section we will show that this is true only for very weak driving fields.

 Out of resonance, $\rho_{00}\simeq 1$ yielding 
 $\langle a \rangle  \simeq -f[(\delta \omega+g)^{-1}+
 (\delta \omega-g)^{-1}]/4$ up to first order in $f$.
As follows from figs.~\ref{fig.2} a) and b) (dashed-dotted red lines), 
the lowest order response $A_0$ coincides with the exact one away from resonance. 
Moreover, $\varphi_0=\pi$ for $\delta \omega \le g$ and  
$\varphi_0=0$ for $\delta \omega > g$. 

At the $N$-photon-resonance, the system can leave  $|\phi_0\rangle$ by tunneling to $|\phi_{N-}\rangle$  and the bath induces decays from $|\phi_{n-}\rangle$ to $|\phi_{n-1 -}\rangle$ along the ladder $N\to N-1 \to ... \to 1 \to 0$. The decay rates (in secular approximation) are 
readily evaluated from Eq.\ (\ref{me}) as  
${\cal L}_{n-,(n-1)-}=(\sqrt{n}+\sqrt{n-1})^2\kappa/4$ for $n\neq 1$ and ${\cal L}_{1-,0}=\kappa/2$. Hence, the rate from the $1$-photon to the $0$-photon state is smallest. 
 Note that the probability of a decay to a state with opposite dressed spin is  small, i.e., 
${\cal L}_{(n-1)-, n+}/{\cal L}_{(n-1)+, n+}\simeq 
1/[16 n \left(n-1/2\right)]$. Hence, the population of the $1$-photon state is always larger than those of the $n$-photon states. 
In addition, depending on the ratio $\kappa/\Omega_N$, qualitatively different stationary populations arise from a competition between 
tunneling from $|\phi_0\rangle$ to the top of the ladder, $|\phi_{N-}\rangle$, and relaxation down the ladder.  
For $\kappa\gg \Omega_N$, the population of $|\phi_0\rangle$ is $\simeq 1$, because damping is more efficient than tunneling. Dissipation then completely washes out the resonance, and the response is the same as that off resonance and thus is in phase with the drive.  
For $\kappa\simeq \Omega_N$, a small population $\rho_{11}$ emerges, contributing $\rho_{11}a_{11}$ with  
$ a_{11} = \frac{f}{2(\varepsilon_{1-}-\varepsilon_{2-})} \frac{3+2\sqrt{2}}{4} + 
\frac{f}{4(\varepsilon_{1-}-\varepsilon_{0})} <0 $
since, in the perturbative regime and for $N<6$, $\varepsilon_{1-}<\varepsilon_{2-},\varepsilon_0$. This leads to a reduced response, forming an antiresonance, see, e.g.,  the $2$-photon antiresonance  shown in Fig.~\ref{fig.1}a. 
For $\kappa\ll \Omega_N$,  tunneling is faster than relaxation and the population of $|\phi_{1-}\rangle$ becomes the largest. Then, the contribution $\rho_{11}a_{11}<0$ dominates, leading to an  out-of-phase oscillation.   
 To be specific, for the $2$-photon resonance of Fig.~\ref{fig.1}, $\kappa/\Omega_{2}=\kappa g (\sqrt{2}f^2)=1.8$ (a), 
$0.29$ (b) and $0.1$ (c).  

\section{Semiclassical approach}
For increasing (but still weak) driving , the response is qualitatively similar, although the perturbative approach becomes inadequate. In fact, as shown in  Fig.~\ref{fig.2}d, at resonance, 
there is a large population $P^*$ of the lowest quasienergy state $|\psi^*\rangle$. As follows from Fig.~\ref{fig.2}c, $|\psi^*\rangle \ne |\phi_{1-}\rangle$. 
In order to account for the importance of $|\psi^*\rangle$, we perform next a semiclassical analysis. We switch to the dressed atom picture, by means of the unitary 
transformation 
\begin{equation}
R=\exp\left({\frac{-3\pi}{8\sqrt{a^\dagger a+\sigma_z/2+1/2}}[a^\dagger \sigma_--a\sigma_+]}\right)\, ,
\end{equation}
 mapping
the JC eigenstates into product states $|n\sigma \rangle$ (with $\sigma=g,e$). Under this transformation,
the undriven JC Hamiltonian becomes diagonal in the spin degrees of freedom and reads 
\begin{eqnarray}
\tilde{H}=|\delta \omega | \left( a^\dagger a+\frac{\sigma_z}{2} +\frac{1}{2}\right)  + g \sigma_z \sqrt{a^\dagger a+\frac{\sigma_z}{2}+\frac{1}{2}} 
\end{eqnarray}
while $\tilde{a}=R^\dagger a R = a+{\cal O}(n^{-1/2})$. This expression follows by expanding the matrix elements of the creation operator in the dressed Fock basis with respect to $n$.
Higher order terms include spin flipping operators. Next, we introduce the canonical 
variables ${\cal X}=\sqrt{\lambda/2}(a^\dagger+a)$ and ${\cal P}=i\sqrt{\lambda/2}(a^\dagger-a)$, where 
$\lambda=\delta \omega^2 / g^2$ is a dimensionless parameter playing the role of $\hbar$. Eventually, neglecting all higher order  terms , we obtain the transformed
 Hamiltonian $\tilde{H}\simeq \delta\omega\lambda^{-1} Q ({\cal X}, {\cal P})$ with 
\begin{equation} \label{hampotfull}
Q ({\cal X}, {\cal P})  =   \frac{{\cal X}^2}{2}+\frac{{\cal P}^2}{2} + \sigma_z  \sqrt{\frac{{\cal X}^2}{2}+\frac{{\cal P}^2}{2}} 
+ \frac{f}{\sqrt{2} g} {\cal X}   \, .
\end{equation}
\begin{figure}
\begin{center}
\includegraphics[width=45mm,keepaspectratio=true]{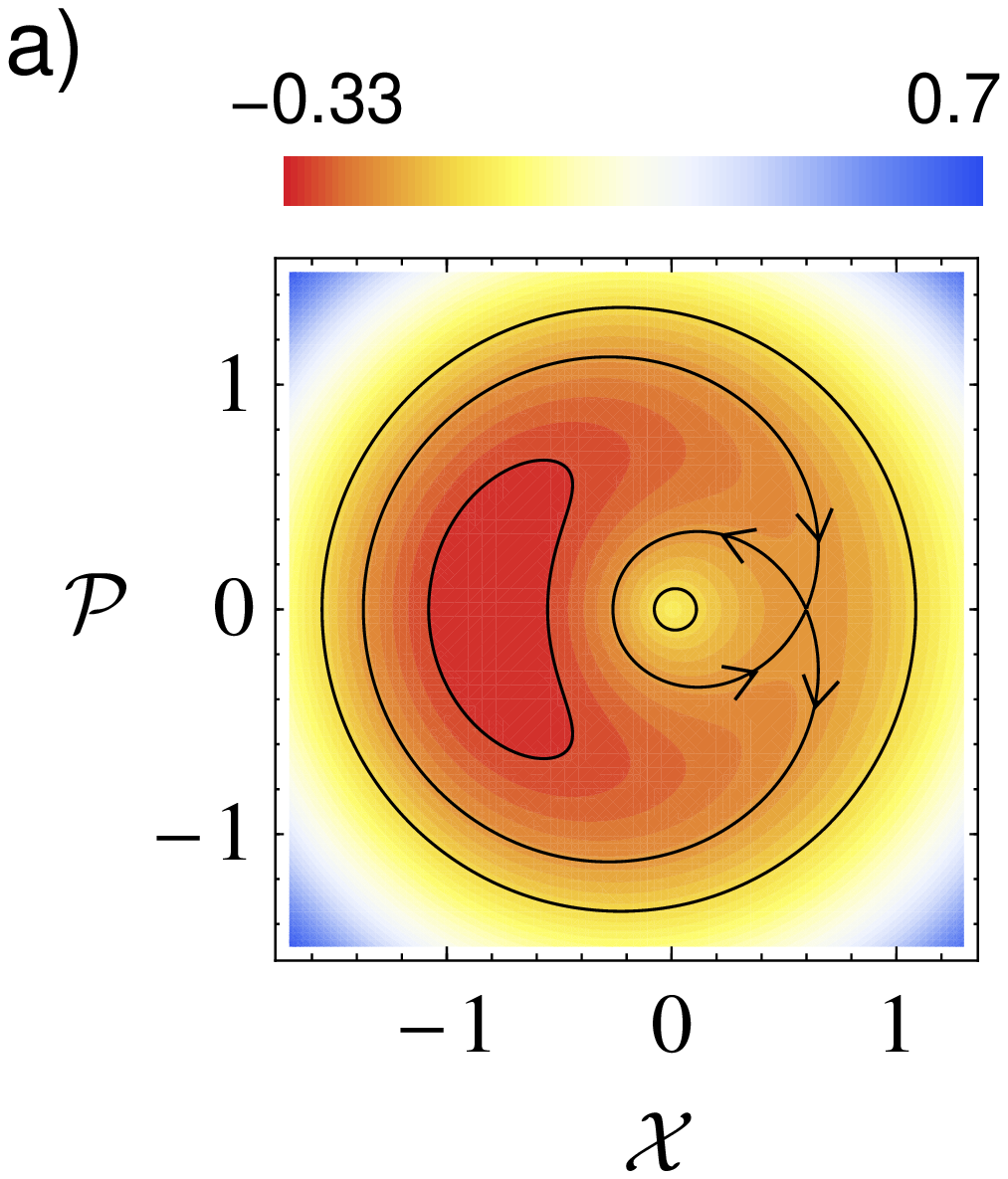}\\
\includegraphics[width=45mm,keepaspectratio=true]{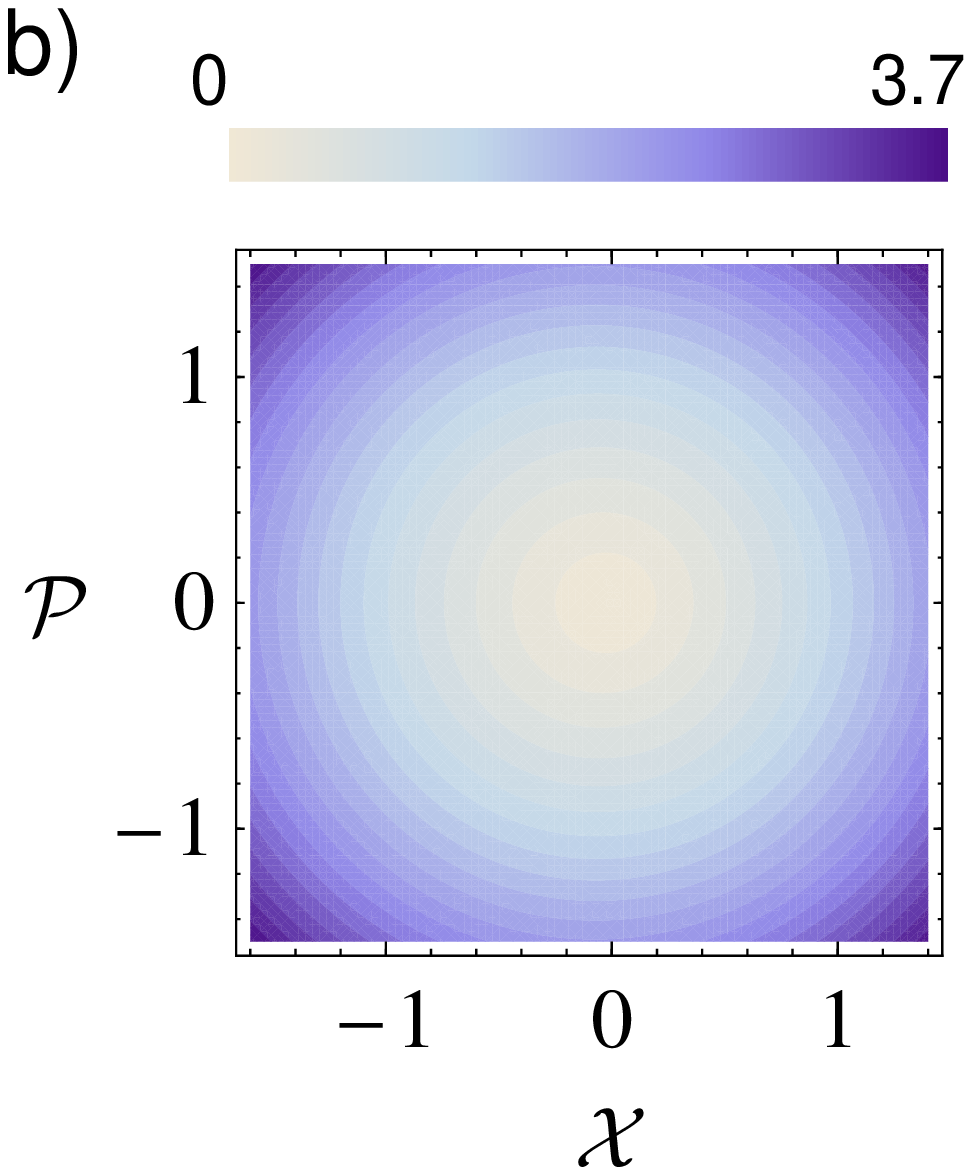}
\caption{\label{fig.3}Contour plot of the quasienergy surfaces $Q({\cal X}, {\cal P})$  for the dressed spin orientation 
$g \, (\sigma_z=-1)$ (a) and $e \, (\sigma_z=+1)$ (b) for $f=0.004 \omega_{\rm r}$ and  $g=0.026\omega_{\rm r}$, yielding  
$f/g=0.154$ and $T_{\rm eff}=0.65$. The solid black lines in (a) indicate quasiclassical orbits and 
include also the separatrix between the two domains of attraction.}
\end{center}
\end{figure}
 It can be interpreted as a quasienergy surface in phase space and is visualised in Fig.~\ref{fig.3}. 
For $\sigma_z=-1$ (Fig.~\ref{fig.3} a), an internal dome appears around 
 ${\cal X} = 0,{\cal P}=0$, such that two quasiclassical orbits encircle  
the inner maximum, one on the external and one on the internal domain. They are degenerate with respect to 
their quasienergies and thereby define a dynamical bistability. Both domains are separated by the 
separatrix shown in Fig.~\ref{fig.3} a as the oriented black solid line. The overall shape of the quasienergy surface 
resembles a Mexican hat. The drive induces a finite tilt of the surface, generating a quasienergy well
around the quasienergy minimum. The orbits near the inner maximum correspond to small photon numbers   
and are not accurately described by the semiclassical approach. On the other hand, the orbits 
on the outer domain correspond to states with large photon numbers. 
Deep in the quantum regime,  discrete multiphoton transitions correspond to tunneling 
transitions between these quasiclassical orbits. The surface for 
the opposite dressed spin orientation $\sigma_z=+1$ is a less
 interesting monotonous function, shown in Fig.~\ref{fig.3} b. 

Let us next consider the dynamics around the minimum located at ${\cal P}_{\rm min}=0$ and 
$ {\cal X}_{\rm min}= -\left( f/g + 1\right)/\sqrt{2}$. A harmonic expansion yields 
the  quasienergy $E^*$ for the state localized at the bottom of the quasienergy surface as
$E^*= \delta\omega\lambda^{-1} [Q({\cal X}_{min},0) +(\lambda/2) \omega^*]$, or written in 
terms of $f$ and $g$ as 
\begin{equation} \label{estar}
E^* =  
-\frac{g^2}{4\delta\omega}\left( \frac{f}{g} + 1\right)^2 +  \frac{\delta\omega}{2} \sqrt{\frac{f}{g+f}} \, , 
\end{equation}
with the effective frequency 
$ \omega^* = \sqrt{f/(g+f)}$. 
This result is correct up to ${\cal O}(\lambda^2)$ and is shown in Fig.~\ref{fig.2}c as orange dashed line. It almost coincides  with the exact result, 
even for small photon numbers $N=1$. 

Up to leading order in $\lambda$, the lowest quasienergy eigenstate
\begin{equation}|\psi^*\rangle=R^{-1} D({\cal X}_{min})S(r)|0\, g\rangle\label{lowestenergystate2}
\end{equation}
is obtained by applying to the vacuum:
i) the squeezing operator $S(r)=\exp \left[r(a^{ 2}-a^{ \dagger2})/2\right]$ with squeeze factor  
$r=-(\ln \omega^*)/2=\ln [1+g/f]/4$, 
ii) the translation $D({\cal X}_{\rm min})=\exp \left[i {\cal P} {\cal X}_{\rm min}/\lambda\right]$   to the  minimum, and iii) $R^{-1}$ to return to the bare atom picture. 
With this, one can compute the Fock-state representation~\cite{yuen} and all expectation values at leading order. 
It turns out that $|\psi^*\rangle$ has sub-Poissonian statistics and shows photon antibunching. This is not surprising, in fact $D({\cal X}_{min})S(r)|0\, g\rangle$ 
is an amplitude squeezed state for $r>0$.

Note that this state $|\psi^*\rangle$ exists for any finite driving strength 
(provided that $\lambda$ be much smaller than the barrier height $f/g$), 
but it is not present in the undriven case. For this reason it is not possible to 
describe it in terms of simple perturbation theory in the driving.
In fact, in the limit of vanishing  driving, the tilt of 
the quasienergy surface vanishes and
 no quasienergy well is developed. 
For weak but finite driving, the well is still very shallow in the momentum direction, 
allowing for large momentum fluctuations of a state confined in it. For increasing driving, the well becomes deeper 
and more symmetric. For  this reason, the squeezing factor decreases for increasing driving. Let us furthermore emphasize 
that  $|\psi^*\rangle$, i.e., the lowest 
quasienergy eigenstate of the driven JC Hamiltonian, can not be identified with the squeezed state investigated in Ref.~\cite{Milburn} (for $\delta\omega= 0$).

The latter  is the eigenstate obtained by starting in the JC groundstate $|\phi_0\rangle$ and by switching on 
adiabatically the driving. It has vanishing quasienergy
for $\delta\omega\to 0$.  Moreover, 
it tends to follow the driving by rotating its spin part and 
squeezing its amplitude fluctuations.
 Hence, as opposed to $|\psi^*\rangle$, its squeezing factor 
{\em increases\/} for increasing driving. We expect the Wigner representation for this state to be 
confined in an area of size $\lambda^2$ around the local quasienergy maximum. 
Unfortunately, it cannot be incorporated in our semiclassical description because 
it corresponds to a small photon number. 
In the following discussion, we will assume that $f\ll g$, so that perturbation theory is still 
valid close to {\em zero\/} quasienergy 
(although it is not valid close to the quasienergy minimum). 
In this case, the JC groundstate $|\phi_0\rangle$ is still an approximate eigenstate of the driven 
JC Hamiltonian.
\begin{figure}
\begin{center}
\includegraphics[width=70mm,keepaspectratio=true]{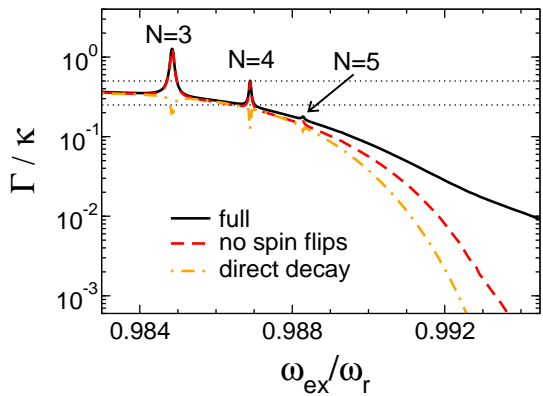}
\caption{\label{fig.4}
Solid line: Smallest eigenvalue of the Lindblad master equation. Dashed red line: 
the same without dissipative spin flips. Dashed-dotted orange line: decay rate for $|\psi^*\rangle \to 
| \phi_{0}^f \rangle$. Dotted lines correspond to $\kappa/2$ and $\kappa/4$. The parameters are 
$g=0.026 \omega_{\rm r}$, $f=0.004\omega_{\rm r}$, and $\kappa=10^{-4} \omega_{\rm r}$.} 
\end{center}
\end{figure}

Next, we consider the dissipative semiclassical dynamics. As will be shown below, 
a separation of time scales exists which defines a fast intrawell and a slow interwell 
relaxation. Deep in the semiclassical limit, the quasienergy states, localized  close to the minimum of 
 one well,  can be obtained as $|\psi^*_n\rangle=R^{-1}b^{\dagger n}R|\psi^*\rangle/\sqrt{n!}$ with 
$b= a\cosh{r}+ a^\dagger\sinh{r}-e^{r}{\cal X}_{\rm min}/\sqrt{2\lambda}$.
In this limit, dissipative transitions occur only between nearest neighbors, with the rates
${\cal L}_{n-1,n}=\kappa n \cosh^2 r$, ${\cal L}_{n,n-1}=\kappa n\sinh^2r$. Here,  the detailed balance condition is fullfilled. 
When the system is initially in a state with a large photon number, it has a large probability to fall in the basin of attraction of the 
 quasipotential minimum  (intrawell relaxation). When also $\kappa\ll g\lambda\omega^*$, detailed balance determines an 
effective Boltzmann distribution $P^*_n=P^* e^{-n \beta_{\rm eff}}$, with 
 effective inverse temperature $\beta_{\rm eff}=2 \ln \coth r$. 
 We emphasize that this link between effective temperature and squeezing can be generalized to any driven quantum system with a smooth 
 quasienergy surface and coupled linearly to a bath (e.g., the Duffing 
 oscillator~\cite{Dykman80,Dmitriev,Dykman89,Peano1,Dykman,Peano2,Peano3,Dykman07} 
or the parametrically driven oscillator~\cite{Dykman06}). 
It can be easily generalized to finite temperatures $T>0$ as well. It turns out that the zero temperature limit applies  when 
$\sinh^2 r$  is much larger than the bosonic occupation number  $\bar{n}(\omega_{\rm ex}/T)$ of the bath at $\omega_{\rm ex}$. 
In the opposite limit, $\beta_{\rm eff}=\omega_{\rm ex}/T$. Since we include here only photon leaking, 
i.e., $\omega_{\rm ex} \gg T$, the effective temperature is still small.   

On the large time scale, the system decays to the $0$-photon state with a forward rate $k_+$ 
for the interwell relaxation. From there, it can return to the basin of attraction of the minimum by a 
driving induced transition with a backward rate $k_-$. 
The stationary populations of the intrawell states (oscillating out of phase)  and the 
$0-$photon state (in phase) are determined by the ratio $k_+/k_-$. 
Off resonance, photon leaking favors the $0-$photon state and $k_+\gg k_-$. 
Approaching a resonance, $k_-$ increases due to an increasing contribution due to tunneling to the escape.  
When the tunneling contribution becomes comparable to $k_+$, the response 
is qualitatively modified yielding an (anti-)resonance.

To render the qualitative discussion more quantitative, we consider the eigenvalues of the Liouville 
operator which governs the Lindblad master equation (\ref{me}). We start deep in the semiclassical regime, i.e., 
for small detuning, where $\omega_{\rm ex}\approx \omega_r$. Here, 
a clear separation of time scales for the dissipative dynamics on the bistable quasienergy surface 
occurs. Well-defined energy wells with a large quasienergy barrier in between exist for small detuning and 
allow for a clear description in terms of a single relaxation rate (see also Ref.\ \cite{rate}). 
In this regime, we find a single eigenvalue $\Gamma$ which 
consists of the sum of $k_-$ and $k_+$, is real and much smaller than the real parts of 
all the other eigenvalues. This is shown in Fig.\ \ref{fig.4} as black solid line. 
For increasing detuning, i.e., 
when $\omega_{\rm ex} \lesssim 0.987 \omega_r$ in Fig.\ \ref{fig.4},  we enter a regime, 
where the separation of time scales 
is not so clearly expressed and $\Gamma$ becomes comparable to the real parts of three 
more eigenvalues. The latter correspond to relaxation (one real eigenvalue) and to decoherence 
(a complex conjugated pair of eigenvalues) involving the pair of states $|\phi_{1+}\rangle$ and 
 $|\phi_0\rangle$. We do not 
show them in the figure, but instead show  the perturbatively determined values 
$\kappa/2$ (relaxation) and $\kappa/4$ (decoherence) out of resonance as dotted horizontal lines. 
The peaks in Fig.\ \ref{fig.4} (black solid lines) are due to  tunneling from the $0$- to $N$-photon state.
When the Rabi frequency $\Omega_N$ is much larger than the $N-$photon dressed state decay rate $\Gamma_N$, 
the system tunnels back and forth many times 
between the two internal domains of the quasienergy surface, before it decays 
to the external domain with rate $k_-\approx \Gamma_N/2$. In the opposite limit $\Gamma_N\gg\Omega_N$, 
the tunneling rate is reduced due to the quasienergy level fluctuations. It becomes 
equal to  $\Omega_N^2/\Gamma_N$ \cite{difftunrate} and coincides approximately with the escape rate
 $k_-$. Away from resonance, $\Gamma\approx k_+$. 

There are three mechanisms of decay from the metastable well: (i) 
The system can decay directly to the $0$-photon state 
with the rate shown in Fig.~\ref{fig.4}, dashed-dotted orange line. 
 (ii) The system can climb up the quasienergy well by 
quantum activation~\cite{Dykman89}. 
Both  associated rates are expected to decrease exponentially, following $\propto e^{-c_i/\delta \omega^2}$, with  some constants 
$c_i$ (the prefactor varies smoothly with $\delta \omega$), which defines the separation of time scales. 
(iii)  For very small detuning, the escape occurs via bath-induced spin flips. In fact, this mechanism is suppressed only as a power-law 
$\Gamma \propto \delta \omega^2$. To separate (ii) from (iii), we show $\Gamma$  without the bath-induced 
spin flips in Fig.~\ref{fig.4} (dashed red line), illustrating that spin flips are dominant when the separation of time scales 
is well defined. Since only a few states close to the bottom are populated, our solution is stable against 
a small dephasing of the oscillator~\cite{Dykman06} or an intrinsic spin relaxation that violates detailed balance. 
The induced spin-flip rate would be small and 
remain finite for $\lambda \to 0$, imposing an upper limit to the lifetime of $|\psi^*\rangle$. 

\section{Perspectives and conclusions}
The transformation of the driven JC Hamiltonian to the quasienergy picture defined by Eq.\ 
(\ref{hampotfull}) allows for a quantitative semiclassical WKB analysis in the spirit of 
Refs.\ \cite{Dykman89,Dykman06}. We defer this to a future publication since it lies beyond the 
scope of this Letter.
Our analysis can be easily extended to any driven nonlinear oscillator coupled bilinearly to a thermal bath. 
 For example, the Duffing oscillator is characterized by two classical stable solutions with 
different amplitudes and opposite oscillation phase. 
The two states can be associated with classical orbits in a quasienergy landscape which also assumes the form 
of a tilted Mexican hat \cite{Dmitriev} like for the JC model. 
Transitions between the two states can be induced by three mechanisms:  (i) 
 thermal activation ``over the barrier'' \cite{Dykman80,Dmitriev}, 
where the thermal fluctuations provide the energy for classical escape, 
 ,  (ii) quantum activation \cite{Dykman89} where bath zero-point quantum fluctuations induce a diffusive motion 
on the quasienergy surface over the barrier, which even at zero temperature can lead to escape, and (iii) 
quantum tunneling\cite{Peano1,Peano2,Peano3}.

For weak driving, the small-oscillation solution can be 
identified with the $0-$photon quantum state. Since it is favored by photon leaking, it has a low effective temperature and can be regarded as 
stable in absence of tunneling. However, 
 this stable state is associated to a relative quasienergy maximum, in contrast to the case of a static bistable potential.  
At a multiphoton resonance, this state becomes metastable and generates an (anti-)resonance of the stationary oscillation and tunneling peaks in the 
interwell relaxation rate as a function of the detuning. 
These features has been already predicted in Refs.~\cite{Peano1,Peano2,Peano3}, but the link 
to the semiclassical picture was not drawn. 

Note that our results on the resonant tunneling transitions in 
interwell dissipative dynamics is not in contradiction with the findings of Ref.~\cite{Dykman89}.
There, it has been shown 
that the  quantum activation rate is always larger than that of dissipative 
tunneling when they are computed semiclassically with logaritmic precision.
Here and in Refs.~\cite{Peano1,Peano2,Peano3}, we show that tunneling is the dominating 
escape mechanism in the complementary {\em deep quantum regime} when it is coherent
 and/or only very few states are confined 
within the quasienergy well (so that logaritmic corrections are relevant).

We furthermore note that the dissipative quantum Duffing 
oscillator could be realized in a Josephson bifurcation amplifier~\cite{jba} 
operating in the quantum regime.

In conclusion, inspired by recent experiments, we have explained the nonlinear response of the 
driven dissipative Jaynes-Cummings model. We have predicted the existence of a metastable squeezed state in the semiclassical limit and drawn a link between effective local temperature and the squeezing parameter. We have analyzed the escape mechanisms from the metastable states and 
found resonant dynamical tunneling. 
Our analysis reveals generic features on the dissipative dynamics of nonlinear driven quantum systems 
for the example of the driven Jaynes-Cummings model, realized in terms of a 
nanocircuit QED setup.

\acknowledgments
We thank M.\ I.\ Dykman, M.\ H.\ Devoret and V.\ Leyton  for useful discussions and 
acknowledge support by the For\-schungs\-f\"or\-der\-ungs\-fonds of the Universit\"at D\"usseldorf, 
 the Excellence Initiative of the German Federal and State Governments, and  the DAAD-PROCOL Program.


\begin{thebibliography}{99}
\bibitem{jaynescummings}
 \Name{Jaynes E. T. \and  Cummings F.W.}
  \REVIEW{Proc. IEEE}{51}{1963}{89}.

\bibitem{Larson}
 \Name{Larson J.}
  \REVIEW{Phys. Scr.}{76}{2007}{146}.

\bibitem{Wallraff} 
\Name{Wallraff A. {\em et al.}}
  \REVIEW{Nature}{431}{2004}{162}.

\bibitem{Chiorescu}
\Name{Chiorescu I. {\em et al.}}
  \REVIEW{Nature}{431}{2004}{159}.

\bibitem{Hatakenaka}
\Name{Hatakenaka N. \and Kurihara S.}
  \REVIEW{Phys. Rev. A}{54}{1996}{1729}.

\bibitem{Winger} 
\Name{Winger M.}
  \REVIEW{Phys. Rev. Lett.}{101}{2008}{226808}.

\bibitem{Bishop} 
\Name{Bishop L. S. {\em et al.}}
  \REVIEW{Nature Phys.}{5}{2009}{105}.

\bibitem{Dykman89} 
\Name{Dykman M. I. \and Smelyanskii V. N.}
  \REVIEW{Sov.\ Phys. JETP}{67}{1988}{1769}.

\bibitem{Dykman06}
\Name{Marthaler M. \and Dykman M. I.}
\REVIEW{Phys. Rev. A}{73}{2006}{042108}.

\bibitem{Dykman07} 
\Name{Dykman M.I.}
  \REVIEW{Phys. Rev. E}{75}{2007}{011101}.

\bibitem{Peano1}
\Name{Peano V. \and Thorwart M.}
\REVIEW{Phys. Rev. B}{70}{2004}{235401}.

\bibitem{Peano2}
\Name{Peano V. \and Thorwart M.}
\REVIEW{Chem. Phys.}{322}{2006}{135}. 

\bibitem{Peano3}  
\Name{Peano V. \and Thorwart M.}
\REVIEW{New J. Phys.}{8}{2006}{21}. 

\bibitem{Dykman80} 
\Name{Dykman M. I. \and Krivoglaz M.A.}
  \REVIEW{Physica A}{104}{1980}{480}.

\bibitem{Dmitriev} 
\Name{Dmitriev A.P. \and D'yakonov M.I.}
  \REVIEW{Sov. Phys. JETP}{63}{1986}{838}.

\bibitem{Dykman}
\Name{Dykman M. I. \and Fistul M.V.}
\REVIEW{Phys. Rev. B}{71}{2005}{140508(R)}.



\bibitem{Houck}  
\Name{Houck  A. A. {\em et al.}}
  \REVIEW{Phys. Rev. Lett.}{101}{2008}{080502}.

\bibitem{Chough}
\Name{Chough Y. T., Nha H. \and An K.}
\REVIEW{J. Phys. Soc. Jpn.}{69}{2000}{4060}.

\bibitem{yuen} 
\Name{Yuen H. P.}
\REVIEW{Phys. Rev. A}{13}{1976}{2226}.

\bibitem{Milburn}\Name{Milburn G. \and Alsing P.}
  \Book{in Directions in Quantum Optics, Lecture Notes in Physics}
  \Editor{Carmichael H. J., Glauber R. J. \and Scully M.\ O.}
  \Vol{561}
  \Publ{Springer, Berlin}
  \Year{2001}
  \Pages{303}{312}. 

\bibitem{rate} 
\Name{H\"anggi P., Talkner P. \and M.\ Borkovec M.}
  \REVIEW{Rev. Mod. Phys.}{62}{1990}{251}.

\bibitem{difftunrate} 
\Name{Dykman M. I. \and Tarasov G. G.}
  \REVIEW{Sov. Phys. JETP}{47}{1978}{557}.

\bibitem{jba} 
\Name{Siddiqi I. {\em et al.}}
  \REVIEW{Phys. Rev. Lett.}{93}{2004}{207002}.

\end{thebibliography}
\end{document}